\PassOptionsToPackage{unicode}{hyperref}
\PassOptionsToPackage{hyphens}{url}
\PassOptionsToPackage{dvipsnames,svgnames,x11names}{xcolor}
\documentclass[
  12pt]{article}

\usepackage{amsmath,amssymb}
\usepackage{iftex}
\ifPDFTeX
  \usepackage[T1]{fontenc}
  \usepackage[utf8]{inputenc}
  \usepackage{textcomp} 
\else 
  \usepackage{unicode-math}
  \defaultfontfeatures{Scale=MatchLowercase}
  \defaultfontfeatures[\rmfamily]{Ligatures=TeX,Scale=1}
\fi
\usepackage{lmodern}
\ifPDFTeX\else  
\fi
\IfFileExists{upquote.sty}{\usepackage{upquote}}{}
\IfFileExists{microtype.sty}{
  \usepackage[]{microtype}
  \UseMicrotypeSet[protrusion]{basicmath} 
}{}
\makeatletter
\@ifundefined{KOMAClassName}{
  \IfFileExists{parskip.sty}{%
    \usepackage{parskip}
  }{
    \setlength{\parindent}{0pt}
    \setlength{\parskip}{6pt plus 2pt minus 1pt}}
}{
  \KOMAoptions{parskip=half}}
\makeatother
\usepackage{xcolor}
\makeatletter
\ifx\paragraph\undefined\else
  \let\oldparagraph\paragraph
  \renewcommand{\paragraph}{
    \@ifstar
      \xxxParagraphStar
      \xxxParagraphNoStar
  }
  \newcommand{\xxxParagraphStar}[1]{\oldparagraph*{#1}\mbox{}}
  \newcommand{\xxxParagraphNoStar}[1]{\oldparagraph{#1}\mbox{}}
\fi
\ifx\subparagraph\undefined\else
  \let\oldsubparagraph\subparagraph
  \renewcommand{\subparagraph}{
    \@ifstar
      \xxxSubParagraphStar
      \xxxSubParagraphNoStar
  }
  \newcommand{\xxxSubParagraphStar}[1]{\oldsubparagraph*{#1}\mbox{}}
  \newcommand{\xxxSubParagraphNoStar}[1]{\oldsubparagraph{#1}\mbox{}}
\fi
\makeatother

\usepackage{longtable,booktabs,array}
\usepackage{calc} 
\usepackage{etoolbox}
\makeatletter
\patchcmd\longtable{\par}{\if@noskipsec\mbox{}\fi\par}{}{}
\makeatother
\IfFileExists{footnotehyper.sty}{\usepackage{footnotehyper}}{\usepackage{footnote}}
\makesavenoteenv{longtable}
\usepackage{graphicx}

\makeatletter
\def\maxwidth{\ifdim\Gin@nat@width>\linewidth\linewidth\else\Gin@nat@width\fi}
\def\maxheight{\ifdim\Gin@nat@height>\textheight\textheight\else\Gin@nat@height\fi}
\makeatother
\setkeys{Gin}{width=\maxwidth,height=\maxheight,keepaspectratio}
\makeatletter
\def\fps@figure{htbp}
\makeatother

\makeatletter
\@ifpackageloaded{caption}{}{\usepackage{caption}}
\AtBeginDocument{%
\ifdefined\contentsname
  \renewcommand*\contentsname{Table of contents}
\else
  \newcommand\contentsname{Table of contents}
\fi
\ifdefined\listfigurename
  \renewcommand*\listfigurename{List of Figures}
\else
  \newcommand\listfigurename{List of Figures}
\fi
\ifdefined\listtablename
  \renewcommand*\listtablename{List of Tables}
\else
  \newcommand\listtablename{List of Tables}
\fi
\ifdefined\figurename
  \renewcommand*\figurename{Figure}
\else
  \newcommand\figurename{Figure}
\fi
\ifdefined\tablename
  \renewcommand*\tablename{Table}
\else
  \newcommand\tablename{Table}
\fi
}
\@ifpackageloaded{float}{}{\usepackage{float}}
\floatstyle{ruled}
\@ifundefined{c@chapter}{\newfloat{codelisting}{h}{lop}}{\newfloat{codelisting}{h}{lop}[chapter]}
\floatname{codelisting}{Listing}

\makeatother
\makeatletter
\@ifpackageloaded{caption}{}{\usepackage{caption}}
\@ifpackageloaded{subcaption}{}{\usepackage{subcaption}}
\makeatother

\ifLuaTeX
  \usepackage{selnolig}  
\fi
\usepackage[]{natbib}
\usepackage{bookmark}

\IfFileExists{xurl.sty}{\usepackage{xurl}}{} 
\hypersetup{
  pdftitle={Title},
  pdfauthor={Author 1; Author 2},
  pdfkeywords={3 to 6 keywords, that do not appear in the title},
  colorlinks=true,
  linkcolor={blue},
  filecolor={Maroon},
  citecolor={Blue},
  urlcolor={Blue},
  pdfcreator={LaTeX via pandoc}}

\newcommand{\anon}{1}

\graphicspath{ {./Figures/} }
\usepackage{algorithm}
\usepackage{algpseudocode}
\algrenewcommand\algorithmicrequire{\textbf{Input:}}
\algrenewcommand\algorithmicensure{\textbf{Output:}}

\usepackage{multirow}
\usepackage{bm}

\begin{document}

\def\spacingset#1{\renewcommand{\baselinestretch}%
{#1}\small\normalsize} \spacingset{1}


\if1\anon
{
  \title{\bf Fisher Scoring for Exact Mat\'{e}rn  Covariance Estimation through Stable Smoothness Optimization}
  \author{Yiping Hong\thanks{
    The authors gratefully acknowledge the King Abdullah University of Science and Technology; National Natural Science Foundation of China (No. 12271286); Hebei Natural Science Foundation (No. A2025105010); and the Academic Initiation Program for Young Scholars at Beijing Institute of Technology (XSQD-6120220282). }\hspace{.2cm}\\
   School of Mathematics and Statistics,  Beijing Institute of Technology \\
   Tangshan Research Institute, Beijing Institute of Technology \\
    and \\
   Sameh Abdulah \\
    Computer, Electrical and Mathematical Sciences and Engineering Division, \\
    King Abdullah University of Science and Technology 
    \\
    and \\
    Marc G. Genton \\
    Computer, Electrical and Mathematical Sciences and Engineering Division, \\
    King Abdullah University of Science and Technology 
    \\
    and \\
    Ying Sun \\
    Computer, Electrical and Mathematical Sciences and Engineering Division, \\
    King Abdullah University of Science and Technology 
    \\
    }
  \maketitle
} \fi

\if0\anon
{
  \bigskip
  \bigskip
  \bigskip
  \begin{center}
    {\LARGE\bf Fisher Scoring for Exact Mat\'{e}rn  Covariance Estimation through Stable Smoothness Optimization}
\end{center}
  \medskip
} \fi

\bigskip
\begin{abstract}
Gaussian Random Fields (GRFs) with Mat\'{e}rn covariance functions have emerged as a powerful framework for modeling spatial processes due to their flexibility in capturing different features of the spatial field. However, the smoothness parameter $\nu$ is challenging to estimate using maximum likelihood estimation (MLE), which involves evaluating the likelihood based on the full covariance matrix of the GRF, due to numerical instability. Moreover, MLE remains computationally prohibitive for large spatial datasets. To address this challenge, we propose the Fisher-BackTracking (Fisher-BT) method, which integrates the Fisher scoring algorithm with a backtracking line search strategy and adopts a series approximation for the modified Bessel function. This method enables an efficient MLE estimation for spatial datasets using the \texttt{ExaGeoStat} high-performance computing framework. Our proposed method not only reduces the number of iterations and accelerates convergence compared to derivative-free optimization methods but also improves the numerical stability of $\nu$ estimation. Through simulations and real-data analysis using a soil moisture dataset covering the Mississippi River Basin, we show that the proposed Fisher-BT method achieves accuracy comparable to existing approaches while significantly outperforming derivative-free algorithms such as BOBYQA and Nelder–Mead in terms of computational efficiency and numerical stability. 

\end{abstract}

\noindent%
\vfill

\newpage
\spacingset{1.8} 
\section{Introduction}

The Gaussian process (GP) model is widely used in spatial statistics and machine learning across diverse fields, such as environmental and earth sciences \citep{abdulah2024boosting}, energy sciences \citep{koncewicz2023fast}, biology \citep{barac2019global}, and forestry \citep{finley2017applying}. Among various covariance structures, the Mat\'{e}rn covariance function stands out in geostatistics for its flexibility in capturing different levels of smoothness in spatial data \citep{stein2012interpolation}. 
However, the smoothness parameter $\nu$ in the Mat\'{e}rn covariance model is usually treated as fixed in practice, primarily due to the computational challenges associated with likelihood-based estimation. For instance, \cite{diggle2007model} pointed out that estimating all three parameters in the Mat\'{e}rn covariance model is very difficult due to the ridges or plateau shape in the log-likelihood surface, so they suggest choosing $\nu$ from a discrete candidate set. In the literature, the most common setting for the smoothness parameter is $\nu=0.5$, which reduces to an exponential covariance model. 

Yet, as demonstrated by \cite{oliveira2022information}, geostatistical data can contain substantial information about the smoothness parameter $\nu$, which reflects the mean square differentiability of the underlying random field.  \cite{hong2021efficiency} showed that inaccurate $\nu$ may lead to inefficient spatial predictions and inaccurate estimations of the prediction error. Therefore, obtaining a numerically stable and precise estimate of $\nu$ is crucial for both inference and prediction. 

To compute the exact maximum likelihood estimator (MLE), meaning that the likelihood is evaluated using the full covariance matrix of the Mat\'{e}rn model, several established software packages are available, including the Fortran program \texttt{MLMATERN} \citep{pardo2009mlmatern} and R packages such as \texttt{fields} \citep{nychka2021fields}, \texttt{geostatsp} \citep{brown2015model}, \texttt{geoR} \citep{ribeiro2003geostatistical}, \texttt{RandomFields} \citep{schlather2015analysis}, and \texttt{ExaGeoStatR} \citep{abdulah2023large}. These packages typically rely on derivative-free methods such as pattern search \citep{hooke1961direct}, Nelder--Mead \citep{nelder1965simplex}, and BOBYQA \citep{powell2009bobyqa}, which overlook the fact that the log-likelihood function of the Matérn covariance model is infinitely differentiable with respect to all parameters. Although derivative-based methods such as BFGS \citep{broyden1970convergence} are used in the \texttt{fields} package, this package does not involve computing the MLE of the smoothness parameter $\nu$. Moreover, the BFGS method does not fully exploit the statistical structure of the likelihood function such as the Fisher information. 

Fisher scoring, a derivative-based optimization method that updates parameters using the expected Fisher information, has been successfully used in approximated MLE settings \citep{geoga2020scalable,guinness2021gaussian}. However, to the best of our knowledge, it has not been employed for exact MLE of the Mat\'{e}rn covariance model. A key obstacle is the numerical instability in computing $\partial_\nu \mathcal{K}_\nu$, where $\mathcal{K}_\nu$ is the modified Bessel function of the second kind of order $\nu$. The finite difference approximations of these derivatives often lead to significant round-off errors and unreliable Hessian estimates \citep{geoga2023fitting}. Although \cite{geoga2023fitting} proposed a stable series approximation for these derivatives, their evaluation for the exact MLE computation was limited to small datasets like $n=512$, where $n$ is the number of spatial observations. 

For larger spatial datasets, the exact MLE computation is limited by the computational complexity of inverting an $n\times n$ covariance matrix, which requires $O(n^3)$ operations and $O(n^2)$ memory. To address this, the \texttt{ExaGeoStat} framework \citep{abdulah2018exageostat,abdulah2023large} leverages state-of-the-art high-performance parallel dense linear algebra libraries, such as \texttt{Chameleon} \citep{agullo2012hybridization}, to accelerate matrix operations in log-likelihood evaluation. This enables exact MLE computations for synthetic and real datasets with up to hundreds of thousands to millions of observations, using parallel and distributed computing. However, its optimization procedure is derivative-free and thus does not exploit the Fisher information structure of the likelihood. In this work, we integrate this framework for exact MLE computation, enabling our proposed derivative-based method to run on large-scale systems. 

In this study, we introduce the Fisher-BackTracking (Fisher-BT) method, a robust and efficient exact MLE computation method for the Mat\'{e}rn covariance model that integrates Fisher scoring with backtracking line search and a Nelder-Mead fallback. By leveraging the high-performance \texttt{ExaGeoStat} framework, our proposed method enables exact inference on large datasets. Our approach leverages series approximations to compute derivatives stably, enabling convergence of the Fisher scoring method and significantly reducing the number of iterations compared to derivative-free optimization methods. Through comprehensive simulations and a real-data analysis of soil moisture over the Mississippi River Basin, we demonstrate that the Fisher-BT achieves competitive accuracy while substantially outperforming existing methods like \texttt{ExaGeoStat}/BOBYQA and \texttt{ExaGeoStat}/Nelder-Mead in computational speed and numerical stability, especially for extreme values of $\nu$. 

The remainder of this article is organized as follows. Section~\ref{Sec:Method} presents the MLE optimization framework and introduces the proposed Fisher-BT algorithm. Section~\ref{Sec:simulations} reports numerical simulation results, while Section~\ref{Sec:applications} demonstrates the method on the real-world soil moisture dataset. Finally, Section~\ref{Sec:conclusions} concludes the paper with a summary and discussion.

\section{Methodology} \label{Sec:Method}


In this section, we present the methodological framework used to compute the exact MLE  under the Mat\'{e}rn covariance model. We first describe the proposed Fisher-BT optimization strategy, which combines Fisher scoring with backtracking line search and a Nelder–Mead fallback to ensure both efficiency and stability. We then detail the computation of the log-likelihood, its gradient, and the Fisher information matrix, including the numerical treatment of derivatives with respect to the smoothness parameter~$\nu$.

\subsection{Fisher-BT Optimization Algorithm}

We consider the calculation of the exact maximum likelihood for the stationary Mat\'{e}rn covariance model. Let $\{Z(\symbfit{s}), \symbfit{s}\in \mathbb{R}^2 \}$ be a zero-mean stationary Gaussian random field, so the covariance between two locations $\symbfit{s}_1, \symbfit{s}_2$ is only related to their distance $\|\symbfit{s}_1 - \symbfit{s}_2\|$. We assume:
\begin{equation*}
	\text{Cov}(Z(\symbfit{s}_1), Z(\symbfit{s}_2)) = C(\|\symbfit{s}_1 - \symbfit{s}_2\|; \symbfit{\theta}), 
\end{equation*}
where, 
\begin{equation*}
	C(h; \symbfit{\theta}) = \frac{\sigma^2}{2^{\nu-1} \Gamma(\nu)} \left( \frac{h}{\alpha} \right)^\nu \mathcal{K}_\nu \left( \frac{h}{\alpha} \right)
\end{equation*}
is the Mat\'{e}rn covariance function with parameters $\symbfit{\theta}=(\sigma^2, \alpha, \nu)^\top$, $\mathcal{K}_\nu$ is the modified Bessel function of the second kind of order $\nu$. Consider the spatial data $\symbfit{Z}=(Z(\symbfit{s}_1), \ldots, Z(\symbfit{s}_n))^\top$ observed on $n$ locations $\symbfit{s}_i, i=1, \ldots, n$, then the joint distribution of $\symbfit{Z}$ is the multivariate Gaussian distribution $N_n(\symbfit{0}, \symbf{\Sigma(\symbfit{\theta})})$, where $[\symbf{\Sigma(\symbfit{\theta})}]_{i,j} = C(\|\symbfit{s}_i - \symbfit{s}_j\|; \symbfit{\theta})$, $i,j=1,\ldots,n$, is the covariance matrix. In this case, the log-likelihood function of $\symbfit{Z}$ is 
\begin{equation} \label{Eq:log_likelihood}
	\ell(\symbfit{\theta}) = -\frac{n}{2}\log(2\pi) - \frac{1}{2}\log\{\det(\symbfit{\Sigma}(\symbfit{\theta}))\} - \frac{1}{2} \symbfit{Z}^\top \symbf{\Sigma}(\symbfit{\theta})^{-1} \symbfit{Z}, 
\end{equation}
and the maximum point of $\ell(\symbfit{\theta})$ is defined as the maximum likelihood estimation (MLE). 


The ExaGeoStat software \citep{abdulah2018exageostat} adopts the BOBYQA algorithm \citep{powell2009bobyqa}, which is a derivative-free algorithm that requires the optimization space to be rectangular in shape. Our motivation is to speed up exact MLE computation using derivative-based methods, such as the Fisher scoring algorithm, to remove the parameter-space constraint, reduce the number of iterations, and thereby accelerate the computation. 


When the smoothness parameter $\nu$ is large, the log-likelihood surface becomes nearly flat in the $\nu$ direction, and the numerical evaluation of the Fisher information matrix becomes unstable, with noticeably fluctuating entries across iterations. In this case, the direct Fisher scoring method may generate overly aggressive updates, destabilizing the optimization procedure. To address this, we propose the Fisher–BT algorithm, which augments Fisher scoring with a backtracking line search and incorporates a Nelder–Mead fallback to enhance robustness. The key idea is to exploit derivative information when it is reliable, and to switch to Nelder--Mead when large values of $\nu$ hinder stable convergence. The overall procedure of the proposed method is summarized in Algorithm~\ref{Alg:Fisher_scoring}.

The algorithm starts by selecting an initial value $\symbfit{\theta}^{(0)}$, as a poor initialization may significantly increase the number of iterations or even prevent convergence of the Fisher scoring procedure. Similarly to the derivative-free optimization algorithm, we first assume that the true parameter $\symbfit{\theta}$ is likely to lie within a cube with the bottom-left vector $\symbfit{\theta}^{(\ell)}=(\sigma_{(\ell)}^2, \alpha_{(\ell)}, \nu_{(\ell)})^\top$ and the upper-right vector $\symbfit{\theta}^{(u)}=(\sigma_{(u)}^2, \alpha_{(u)}, \nu_{(u)})^\top$. Then we consider nine candidates for the initial values of the L9 design provided in Table \ref{tab:L9_design} and choose the parameter with the highest log-likelihood value as the initial value $\symbfit{\theta}^{(0)}$. Note that unlike the \texttt{ExaGeoStat}/BOBYQA algorithm, the cube $\Theta = [\sigma_{(\ell)}^2, \sigma_{(u)}^2]\times [\alpha_{(\ell)}, \alpha_{(u)}]\times [\nu_{(\ell)}, \nu_{(u)}]$ is only related to the initial value selection, so the final MLE result could be outside of this cube, which may occur when the true value of $\nu$ is large in our numerical experiments introduced in Section \ref{Sec:simulations}. 

\begin{algorithm}[htbp]
	\caption{Our proposed Fisher-BT algorithm}
	\label{Alg:Fisher_scoring}
	\begin{algorithmic}
		\Require The initial guess vectors $\symbfit{\theta}^{(\ell)}$ and $\symbfit{\theta}^{(u)}$; 
		\Ensure The maximum likelihood estimation $\hat{\symbfit{\theta}}_{MLE}$ and its Fisher information matrix $I(\hat{\symbfit{\theta}}_{MLE})$. 
		\State Use L9 design to determine the initial parameter $\symbfit{\theta}^{(0)}$; 
		\State Set $t=0$;
		\State Perform one Fisher scoring iteration: compute $\ell(\symbfit{\theta}^{(0)})$, $\nabla \ell(\symbfit{\theta}^{(0)})$, $\symbfit{I}(\symbfit{\theta}^{(0)})$, then compute $\symbfit{\phi}^{(0)} \gets \symbfit{I}(\symbfit{\theta}^{(0)})^{-1} \nabla \ell(\symbfit{\theta}^{(0)})$; 
		\While{the stopping condition and the shift condition is not met} 
		\State Check Armijo condition for $\symbfit{\phi}^{(t)}$; 
		\While{The Armijo condition is not met}
		\State $\symbfit{\phi}^{(t)} \gets \symbfit{\phi}^{(t)}/2$; 
		\State Check Armijo condition for the updated $\symbfit{\phi}^{(t)}$; 
		\EndWhile
		\State $\symbfit{\theta}^{(t+1)} \gets \symbfit{\theta}^{(t)} + \symbfit{\phi}^{(t)}$; 
		\State $t \gets t+1$; 
		\State Perform one Fisher scoring iteration: compute $\ell(\symbfit{\theta}^{(t)})$, $\nabla \ell(\symbfit{\theta}^{(t)})$, $\symbfit{I}(\symbfit{\theta}^{(t)})$, then compute $\symbfit{\phi}^{(t)} \gets \symbfit{I}(\symbfit{\theta}^{(t)})^{-1} \nabla \ell(\symbfit{\theta}^{(t)})$; 
		\EndWhile
		\If{The stopping condition is not met}
		\State Perform Nelder-Mead optimization with the initial value as $\symbfit{\theta}^{(t)}$ and the same stopping condition; 
		\State Update $\symbfit{\theta}^{(t)}$ to the MLE result in the previous step; 
		\State Compute $\symbfit{I}(\symbfit{\theta}^{(t)})$ as the Fisher information matrix. 
		\EndIf
		\State $\hat{\symbfit{\theta}}_{MLE} \gets \symbfit{\theta}^{(t)}$, $\symbfit{I}(\hat{\symbfit{\theta}})_{MLE} \gets \symbfit{I}(\symbfit{\theta}^{(t)})$. 
	\end{algorithmic}
\end{algorithm}

\begin{table}[htbp]
	\centering
	\caption{The initial value candidates for the Fisher scoring algorithm}
	\begin{tabular}{cccc}
		\hline
		No. & $\theta_1$ & $\theta_2$ & $\theta_3$ \\ \hline
		1 & $\frac{1}{2}\sigma_{(\ell)}^2 + \frac{1}{2} \sigma_{(u)}^2$ & $\frac{1}{2}\alpha_{(\ell)}+\frac{1}{2}\alpha_{(u)}$ & $\frac{1}{2}\nu_{(\ell)}+\frac{1}{2}\nu_{(u)}$ \\
		2 & $\frac{1}{2}\sigma_{(\ell)}^2 + \frac{1}{2} \sigma_{(u)}^2$ & $\frac{5}{6}\alpha_{(\ell)}+\frac{1}{6}\alpha_{(u)}$ & $\frac{5}{6}\nu_{(\ell)}+\frac{1}{6}\nu_{(u)}$ \\
		3 & $\frac{1}{2}\sigma_{(\ell)}^2 + \frac{1}{2} \sigma_{(u)}^2$ & $\frac{1}{6}\alpha_{(\ell)}+\frac{5}{6}\alpha_{(u)}$ & $\frac{1}{6}\nu_{(\ell)}+\frac{5}{6}\nu_{(u)}$ \\
		4 & $\frac{5}{6}\sigma_{(\ell)}^2 + \frac{1}{6} \sigma_{(u)}^2$ & $\frac{1}{2}\alpha_{(\ell)}+\frac{1}{2}\alpha_{(u)}$ & $\frac{5}{6}\nu_{(\ell)}+\frac{1}{6}\nu_{(u)}$ \\
		5 & $\frac{5}{6}\sigma_{(\ell)}^2 + \frac{1}{6} \sigma_{(u)}^2$ & $\frac{5}{6}\alpha_{(\ell)}+\frac{1}{6}\alpha_{(u)}$ & $\frac{1}{6}\nu_{(\ell)}+\frac{5}{6}\nu_{(u)}$ \\
		6 & $\frac{5}{6}\sigma_{(\ell)}^2 + \frac{1}{6} \sigma_{(u)}^2$ & $\frac{1}{6}\alpha_{(\ell)}+\frac{5}{6}\alpha_{(u)}$ & $\frac{1}{2}\nu_{(\ell)}+\frac{1}{2}\nu_{(u)}$ \\
		7 & $\frac{1}{6}\sigma_{(\ell)}^2 + \frac{5}{6} \sigma_{(u)}^2$ & $\frac{1}{2}\alpha_{(\ell)}+\frac{1}{2}\alpha_{(u)}$ & $\frac{1}{6}\nu_{(\ell)}+\frac{5}{6}\nu_{(u)}$ \\
		8 & $\frac{1}{6}\sigma_{(\ell)}^2 + \frac{5}{6} \sigma_{(u)}^2$ & $\frac{5}{6}\alpha_{(\ell)}+\frac{1}{6}\alpha_{(u)}$ & $\frac{1}{2}\nu_{(\ell)}+\frac{1}{2}\nu_{(u)}$ \\
		9 & $\frac{1}{6}\sigma_{(\ell)}^2 + \frac{5}{6} \sigma_{(u)}^2$ & $\frac{1}{6}\alpha_{(\ell)}+\frac{5}{6}\alpha_{(u)}$ & $\frac{5}{6}\nu_{(\ell)}+\frac{1}{6}\nu_{(u)}$ \\ \hline
	\end{tabular}
	\label{tab:L9_design}
\end{table}

After selecting the initial value, the algorithm starts the Fisher scoring algorithm using a backtracking line search. To obtain the increment vector $\symbfit{\phi}^{(t)}$, we need not only to compute the log-likelihood function $\ell(\symbfit{\theta})$, but also to compute its gradient and the Fisher information matrix. 
Denote $\symbfit{\theta}=(\theta_1, \theta_2, \theta_3)^\top$, then the partial derivatives are computed by 
\begin{equation} \label{Eq:derivative_l}
	\ell_i(\symbfit{\theta}) := \frac{\partial \ell(\symbfit{\theta})}{\partial \theta_i} = -\frac{1}{2} \text{trace}(\symbf{\Sigma}^{-1} \symbf{\Sigma}_i) + \frac{1}{2} \symbfit{Z}^\top \symbf{\Sigma}^{-1} \symbf{\Sigma}_i \symbf{\Sigma}^{-1} \symbfit{Z}, 
\end{equation} 
and the entries of $\symbfit{I}(\symbfit{\theta})$ are computed by
\begin{equation} \label{Eq:Fisher_l}
	[\symbfit{I}(\symbfit{\theta})]_{i, j} = \text{E}_{\symbfit{\theta}}\left(-\frac{\partial^2 \ell(\symbfit{\theta})}{\partial \theta_i \partial \theta_j} \right) = \frac{1}{2} \text{trace} (\symbf{\Sigma}^{-1} \symbf{\Sigma}_i \symbf{\Sigma}^{-1} \symbf{\Sigma}_j), 
\end{equation}
where $[\symbf{\Sigma}_i]_{j, k} = \frac{\partial}{\partial \theta_i} C(\|\symbfit{s}_j - \symbfit{s}_k\|; \symbfit{\theta})$ is the element-wise partial derivative of the covariance matrix with respect to $\theta_i$. The detailed algorithm for computing $\ell(\symbfit{\theta})$, $\nabla \ell(\symbfit{\theta})$, and $\symbfit{I}(\symbfit{\theta})$ will be introduced in Section \ref{Sec:sub_derivative_alg}, in which the terms involving derivatives with respect to $\nu$ are computed using a series approximation algorithm similar to \cite{geoga2023fitting}. 

The Armijo condition ensures that a step size $s$ satisfies 
\begin{equation} \label{Eq:Armijo_def}
	\ell(\symbfit{\theta}^{(t)}+s \symbfit{\phi}^{(t)}) \ge \ell(\symbfit{\theta}^{(t)}) + c s \nabla \ell(\symbfit{\theta}^{(t)}) \cdot \symbfit{\phi}^{(t)}, 
\end{equation}
where $c$ is a small positive constant. This condition requires the optimization process to make sufficient progress in each iteration. The backtracking line search method checks $s=1$ for this condition and shrinks $s$ to $\rho s$ for a certain $0<\rho<1$ until this condition is met. In our procedure, we choose $c=0.001$, $\rho=0.5$. We also relax the Armijo condition, so the condition is considered satisfied when the left-hand side minus the right-hand side in \eqref{Eq:Armijo_def} is greater than $-0.001$. This can prevent the step size from becoming too small for large values of $\nu$. Thus, the Armijo condition in Algorithm \ref{Alg:Fisher_scoring} is 
\begin{equation*}
	\ell(\symbfit{\theta}^{(t)}+\symbfit{\phi}^{(t)}) \ge \ell(\symbfit{\theta}^{(t)}) + 0.001 \nabla \ell(\symbfit{\theta}^{(t)}) \cdot \symbfit{\phi}^{(t)} - 0.001, 
\end{equation*}
and $\symbfit{\phi}^{(t)}$ is updated to $\symbfit{\phi}^{(t)}/2$ until this condition is met. After this condition is satisfied, the next input parameter for Fisher scoring is $\symbfit{\theta}^{(t+1)} = \symbfit{\theta}^{(t)}+\symbfit{\phi}^{(t)}$, until the stopping condition or the shift condition is met. 

We choose the stopping condition as $\|\nabla \ell(\symbfit{\theta}^{(t)})\|_2 \le 0.001$. When this condition holds, the Fisher scoring method converges. However, in some cases, such as when $\nu$ is too large or too small, the Fisher scoring method may fail to converge. Therefore, we set the shift condition using the number of calls for the log-likelihood function and its derivatives. When the program tries to compute $\ell(\symbfit{\theta})$ more than 60 times (including finding the initial value) or tries to compute $\nabla \ell(\symbfit{\theta})$ more than 20 times, the shift condition is met, meaning that the change of the optimization method is preferred in this case. Thus, we shift to the Nelder-Mead optimization, using the last parameter value $\symbfit{\theta}^{(t)}$ as the initial value when the stopping condition is not met and the shift condition is met. In the Nelder-Mead method, we consider that the stopping condition is satisfied when an optimization step changes the value of $\ell(\symbfit{\theta})$ by less than $10^{-9}$. The Nelder-Mead method computes the final estimate $\symbfit{\theta}^{(t)}$, and outputs the Fisher information matrix $\symbfit{I}(\symbfit{\theta}^{(t)})$ after one Fisher matrix computation. 

\subsection{Likelihood Computation Algorithm}
\label{Sec:sub_derivative_alg}

In this section, we present detailed algorithms for computing the log-likelihood function and its derivatives. The algorithm for computing $\ell(\symbfit{\theta})$ is provided in Algorithm \ref{Alg:Log_lik}, which mainly involves Cholesky factorization. 
To compute $\nabla \ell(\symbfit{\theta})$ and $\symbfit{I}(\symbfit{\theta})$ by \eqref{Eq:derivative_l} and \eqref{Eq:Fisher_l}, we need to deal with $\symbf{\Sigma}_i$ and $\symbf{\Sigma}^{-1}\symbf{\Sigma}_i$, which are the most time consuming parts of the computation. 
For derivative matrices $\symbf{\Sigma}_i$, note that $\symbf{\Sigma}_{\sigma^2} = \frac{1}{\sigma^2} \symbf{\Sigma}$. $\symbf{\Sigma}_\alpha$ is computed directly using 
\begin{equation*}
	\frac{\partial }{\partial \alpha} C(h; \symbfit{\theta}) = \frac{1}{2^{\nu-1} \Gamma(\nu)} \cdot \frac{\sigma^2}{\alpha}\cdot \left( \frac{h}{\alpha} \right)^{\nu+1} {\cal K}_{\nu-1} \left( \frac{h}{\alpha} \right). 
\end{equation*}

\begin{algorithm}[htbp]
	\caption{Computing the log-likelihood}
	\label{Alg:Log_lik}
	\begin{algorithmic}
		\Require The parameter $\symbfit{\theta}$; Spatial data $\symbfit{Z}$ observed on locations $\symbfit{s}_i$, $i=1, \ldots, n$
		\Ensure Log-likelihood function $\ell(\symbfit{\theta})$. 
		\State Compute the covariance matrix $\symbf{\Sigma} = \symbf{\Sigma}(\symbfit{\theta})$. 
		\State Perform Cholesky factorization $\symbf{\Sigma} = \symbfit{L}\symbfit{L}^\top$, where $\symbfit{L}$ is the lower-triangle matrix. 
		\State Compute $\log\left\{ \det (\symbf{\Sigma}) \right\} = 2 \cdot \sum_{i=1}^n \log(\ell_{ii})$, where $\ell_{ii}$ is the $i$-th diagonal term of $\symbfit{L}$. 
		\State Compute $\symbfit{Y} = \symbfit{L}^{-1}\symbfit{Z}$. 
		\State Compute $\symbfit{Z}^\top \symbfit{\Sigma}^{-1} \symbfit{Z} = \symbfit{Y}^\top \symbfit{Y}$. 
		\State Compute $\ell(\symbfit{\theta})$ by \eqref{Eq:log_likelihood} and the preceding results. 
	\end{algorithmic}
\end{algorithm}

The derivative matrix $\symbf{\Sigma}_\nu$ is computed from the derivative of $x^\nu {\cal K}_\nu(x)$ using the series approximation algorithm proposed by \cite{geoga2023fitting}. 
The detailed procedure for evaluating this derivative is provided in the Supplementary Material. 
To calculate $\mathrm{trace}(\symbf{\Sigma}^{-1}\symbf{\Sigma}_i \symbf{\Sigma}^{-1} \symbf{\Sigma}_j)$, we employ the following trace identity. 
Let $\symbfit{A}$ and $\symbfit{B}$ be real $n \times n$ matrices; then,

\begin{equation} \label{Eq:Frob_norm}
	\text{trace}(\symbfit{AB}) = \frac{1}{2}\left( \|\symbfit{A}+\symbfit{B}^\top \|_F^2 - \|\symbfit{A}\|_F^2 - \|\symbfit{B}\|_F^2 \right), 
\end{equation}
where $\|\cdot \|_F^2$ is the Frobenius norm, taking only $O(n^2)$ computations. The algorithm for computing $\ell(\symbfit{\theta})$, $\nabla \ell(\symbfit{\theta})$, and $\symbfit{I}(\symbfit{\theta})$ in a row is introduced in Algorithm \ref{Alg:Log_lik_deriv}. 

\begin{algorithm}[htbp]
	\caption{Computing the log-likelihood, gradient, and Fisher information matrix}
	\label{Alg:Log_lik_deriv}
	\begin{algorithmic}
		\Require The parameter $\symbfit{\theta}$; Spatial data $\symbfit{Z}$ observed on locations $\symbfit{s}_i$, $i=1, \ldots, n$. 
		\Ensure Log-likelihood $\ell(\symbfit{\theta})$, its gradient $\nabla \ell(\symbfit{\theta})$, and the Fisher information $\symbfit{I}(\symbfit{\theta})$. 
		\State \textbf{Step 1: Compute $\ell(\symbfit{\theta})$. }
		\State Compute the covariance matrix $\symbf{\Sigma} = \symbf{\Sigma}(\symbfit{\theta})$. 
		\State Perform Cholesky factorization $\symbf{\Sigma} = \symbfit{L}\symbfit{L}^\top$, where $\symbfit{L}$ is the lower-triangle matrix. 
		\State Compute $\log\left\{ \det (\symbf{\Sigma}) \right\} = 2 \cdot \sum_{i=1}^n \log(\ell_{ii})$, where $\ell_{ii}$ is the $i$-th diagonal term of $\symbfit{L}$. 
		\State Compute $\symbfit{Y} = \symbfit{L}^{-1}\symbfit{Z}$ and $\symbfit{Z}^\top \symbf{\Sigma}^{-1} \symbfit{Z} = \symbfit{Y}^\top \symbfit{Y}$. 
		\State Compute $l(\symbfit{\theta})$ by \eqref{Eq:log_likelihood} and the preceding results. 
		\State \textbf{Step 2: Compute $\ell_1(\symbfit{\theta})$ and $[\symbfit{I}(\symbfit{\theta})]_{1, 1}$. }
		\State Compute $\ell_1(\symbfit{\theta}) = \frac{1}{2 \sigma^2}(\symbfit{Z}^\top \symbf{\Sigma}^{-1} \symbfit{Z} - n)$ and $[\symbfit{I}(\symbfit{\theta})]_{1, 1} = \frac{n}{2(\sigma^2)^2}$. 
		\State \textbf{Step 3: Compute $\ell_2(\symbfit{\theta})$ and $[\symbfit{I}(\symbfit{\theta})]_{j, k}$, where $j, k \le 2$. }
		\State Compute the derivative covariance matrix $\symbf{\Sigma}_\alpha = \frac{\partial}{\partial \alpha} \symbf{\Sigma}(\symbfit{\theta})$. 
		\State Compute $\symbfit{A} \gets \symbf{\Sigma}^{-1} \symbf{\Sigma}_\alpha$ and its trace, $\text{trace}(\symbf{\Sigma}^{-1} \symbf{\Sigma}_\alpha)$. 
		\State Compute $[\symbfit{I}(\symbfit{\theta})]_{1, 2} = [\symbfit{I}(\symbfit{\theta})]_{2, 1} = \frac{1}{2\sigma^2} \text{trace}(\symbf{\Sigma}^{-1} \symbf{\Sigma}_\alpha)$. 
		\State Compute vectors $\symbfit{w} \gets \symbf{\Sigma}^{-1}\symbfit{Z}$ and $\symbfit{v} \gets \symbf{\Sigma}_\alpha \symbf{\Sigma}^{-1}\symbfit{Z}$, and $\symbfit{w}^\top \symbfit{v} = \symbfit{Z}^\top \symbf{\Sigma}^{-1} \symbf{\Sigma}_\alpha \symbf{\Sigma}^{-1}\symbfit{Z}$. 
		\State Compute $\ell_2(\symbfit{\theta})$ by \eqref{Eq:derivative_l}, $\symbfit{w}^\top \symbfit{v}$, and $\text{trace}(\symbf{\Sigma}^{-1} \symbf{\Sigma}_\alpha)$. 
		\State Compute $[\symbfit{I}(\symbfit{\theta})]_{2, 2}$ by \eqref{Eq:Fisher_l} and \eqref{Eq:Frob_norm}, where $\symbfit{A} = \symbfit{B} = \symbf{\Sigma}^{-1} \symbf{\Sigma}_\alpha$. 
		\State \textbf{Step 4: Compute $\ell_3(\symbfit{\theta})$ and $[\symbfit{I}(\symbfit{\theta})]_{j, k}$, where $j, k \le 3$. }
		\State Compute the derivative covariance matrix $\symbf{\Sigma}_\nu = \frac{\partial}{\partial \nu} \symbf{\Sigma}(\symbfit{\theta})$. 
		\State Compute $\symbfit{A} \gets \symbf{\Sigma}^{-1} \symbf{\Sigma}_\nu$ and its trace, $\text{trace}(\symbf{\Sigma}^{-1} \symbf{\Sigma}_\nu)$. 
		\State Compute $[\symbfit{I}(\symbfit{\theta})]_{1, 3} = [\symbfit{I}(\symbfit{\theta})]_{3, 1} = \frac{1}{2\sigma^2} \text{trace}(\symbf{\Sigma}^{-1} \symbf{\Sigma}_\nu)$. 
		\State Compute vectors $\symbfit{w} \gets \symbf{\Sigma}^{-1}\symbfit{Z}$ and $\symbfit{v} \gets \symbf{\Sigma}_\nu \symbf{\Sigma}^{-1}\symbfit{Z}$, and $\symbfit{w}^\top \symbfit{v} = \symbfit{Z}^\top \symbf{\Sigma}^{-1} \symbf{\Sigma}_\nu \symbf{\Sigma}^{-1}\symbfit{Z}$. 
		\State Compute $\ell_3(\symbfit{\theta})$ by \eqref{Eq:derivative_l}, $\symbfit{w}^\top \symbfit{v}$, and $\text{trace}(\symbf{\Sigma}^{-1} \symbf{\Sigma}_\nu)$. 
		\State Compute $[\symbfit{I}(\symbfit{\theta})]_{2, 3} = [\symbfit{I}(\symbfit{\theta})]_{3, 2}$ by \eqref{Eq:Fisher_l} and \eqref{Eq:Frob_norm}, where $\symbfit{A} = \symbf{\Sigma}^{-1} \symbf{\Sigma}_\alpha$, $\symbfit{B} = \symbf{\Sigma}^{-1} \symbf{\Sigma}_\nu$. 
		\State Compute $[\symbfit{I}(\symbfit{\theta})]_{3, 3}$ by \eqref{Eq:Fisher_l} and \eqref{Eq:Frob_norm}, where $\symbfit{A}=\symbfit{B} = \symbf{\Sigma}^{-1} \symbf{\Sigma}_\nu$. 
	\end{algorithmic}
\end{algorithm}


\section{Simulations}
\label{Sec:simulations}

This section demonstrates the computational efficiency and estimation accuracy of the proposed Fisher-BT method, compared with the ExaGeoStat framework employing the BOBYQA and Nelder–Mead optimization algorithms, denoted as \texttt{ExaGeoStat}/BOBYQA and \texttt{ExaGeoStat}/Nelder-Mead, respectively. The data are generated from a stationary Gaussian random field with a Mat\'{e}rn covariance function, where the true parameter values $(\sigma^2, \alpha, \nu)$ are set to $(1, 0.1, 0.5)$, $(0.1, 0.1, 0.1)$, $(0.05, 0.05, 0.05)$, $(2, 0.8, 1)$, and $(1.5, 1.55, 1.3)$. These settings involve a case with a moderate value of $\nu$, two relatively small cases of $\nu$, and two relatively large cases of $\nu$. We set the data size as $n=1{,}600$, $3{,}600$, and $6{,}400$. For each sample size $n$ and true parameter configuration, we generate $M = 50$ random realizations from the Mat\'{e}rn model and compute the exact maximum likelihood estimates (MLEs) using the three optimization methods described earlier. 
In the \texttt{ExaGeoStat}/BOBYQA method, the optimization ranges are $(\sigma^2, \alpha, \nu) \in [0.01, 5] \times [0.01, 5] \times [0.01, 2]$, 
whereas in the \texttt{ExaGeoStat}/Nelder-Mead method, the initial optimization value is set to the midpoint of these ranges, $(2.505, 2.505, 1.005)$. 
For the Fisher-BT method, the initial values are selected as shown in Table~\ref{tab:L9_design}, where $\symbfit{\theta}^{(\ell)} = (0.01, 0.01, 0.01), \symbfit{\theta}^{(u)}=(5, 5, 2)$ are vertices of the \texttt{ExaGeoStat}/BOBYQA optimization range, ensuring the fairness of the initial value selection.

\begin{figure}[htbp]
	\centering
	\begin{subfigure}[b]{0.75\textwidth}
		\centering
		\includegraphics[width=\textwidth]{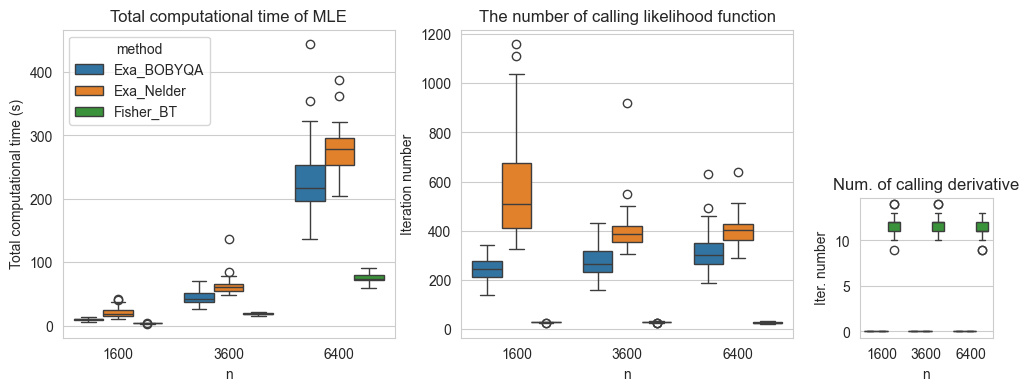}
		\caption{$(\sigma^2, \alpha, \nu) = (1.0, 0.1, 0.5)$}
		\label{subfig:time_10105}
	\end{subfigure}
	\hfill
	\begin{subfigure}[b]{0.75\textwidth}
		\centering
		\includegraphics[width=\textwidth]{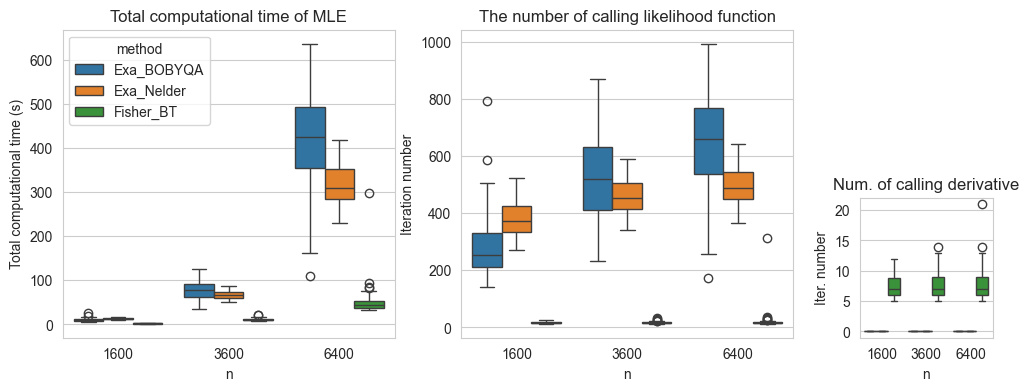}
		\caption{$(\sigma^2, \alpha, \nu) = (2.0, 0.8, 1.0)$}
		\label{subfig:time_20810}
	\end{subfigure}
	\hfill
	\begin{subfigure}[b]{0.75\textwidth}
		\centering
		\includegraphics[width=\textwidth]{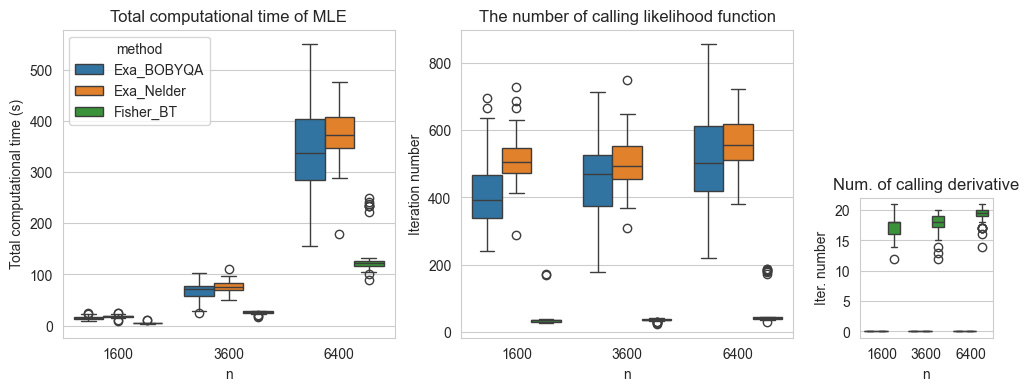}
		\caption{$(\sigma^2, \alpha, \nu) = (0.1, 0.1, 0.1)$}
		\label{subfig:time_010101}
	\end{subfigure}
	\hfill
	\caption{Computational time, the number of iterations, and the number of calling the derivative functions for different MLE algorithms for moderate true values of $\nu$. }
	\label{fig:time1}
\end{figure}

\begin{figure}[htb]
	\centering
	\begin{subfigure}[b]{0.75\textwidth}
		\centering
		\includegraphics[width=\textwidth]{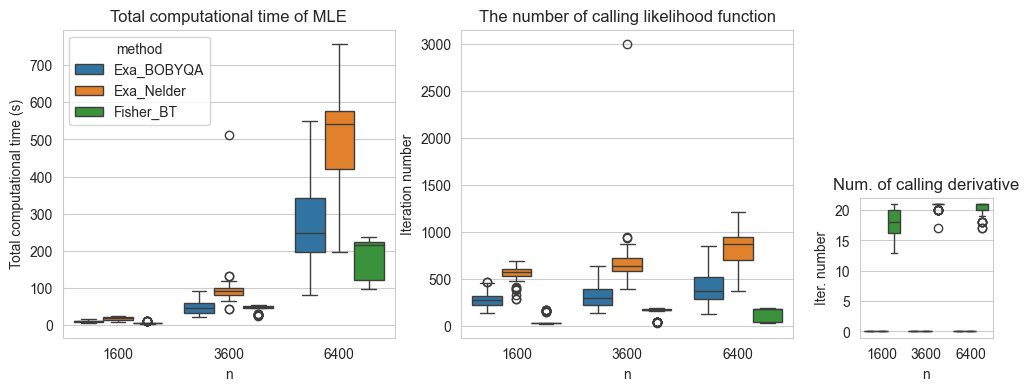}
		\caption{$(\sigma^2, \alpha, \nu) = (0.05, 0.05, 0.05)$}
		\label{subfig:time_005005005}
	\end{subfigure}
	\hfill
	\begin{subfigure}[b]{0.75\textwidth}
		\centering
		\includegraphics[width=\textwidth]{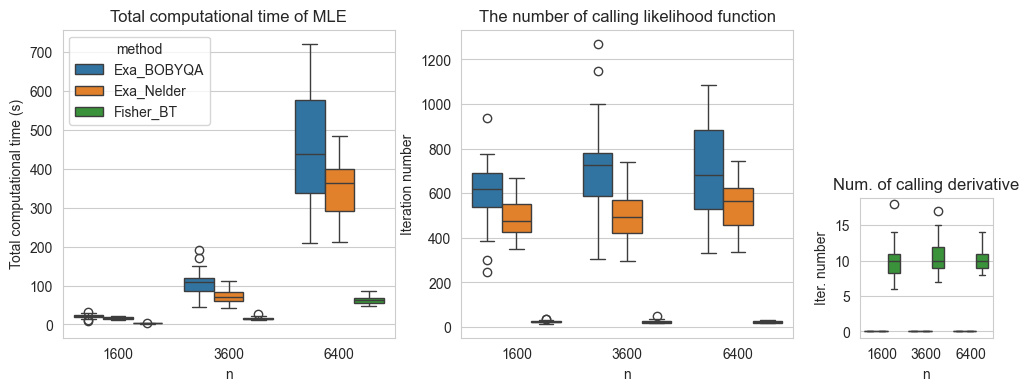}
		\caption{$(\sigma^2, \alpha, \nu) = (1.5, 1.55, 1.3)$}
		\label{subfig:time_1515513}
	\end{subfigure}
	\caption{Computational time, the number of iterations, and the number of calls to the derivative functions for different MLE algorithms for small and large true values of $\nu$. }
	\label{fig:time2}
\end{figure}

First, we present the computational time and the number of iterations required for convergence for the three methods in Figures~\ref{fig:time1} and~\ref{fig:time2}. As shown, the proposed Fisher-BT algorithm offers a significant computational advantage over ExaGeoStat for both optimization algorithms. The only exception is $n=3{,}600$, $\nu=0.05$, in which case the proposed algorithm still has comparable computational performance to the \texttt{ExaGeoStat}/BOBYQA method. 
For the other two methods, the \texttt{ExaGeoStat}/BOBYQA method is faster with small or moderate $\nu$, whereas the \texttt{ExaGeoStat}/Nelder-Mead method is faster for larger $\nu$, such as $\nu=1.0$ or $1.3$. 
Note that the BOBYQA method is the default optimization method in the ExaGeoStat optimization framework in \cite{abdulah2018exageostat}.

\begin{figure}[htbp]
	\centering
	\begin{subfigure}[b]{0.3\textwidth}
		\centering
		\includegraphics[width=\textwidth]{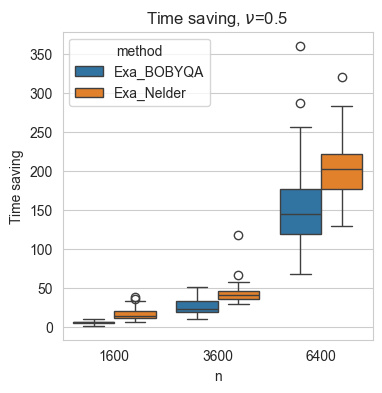}
		\caption{$\nu=0.5$}
	\end{subfigure}
	\hfill
	\begin{subfigure}[b]{0.3\textwidth}
		\centering
		\includegraphics[width=\textwidth]{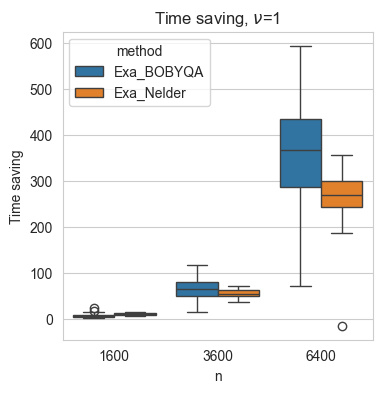}
		\caption{$\nu=1.0$}
	\end{subfigure}
	\hfill
	\begin{subfigure}[b]{0.3\textwidth}
		\centering
		\includegraphics[width=\textwidth]{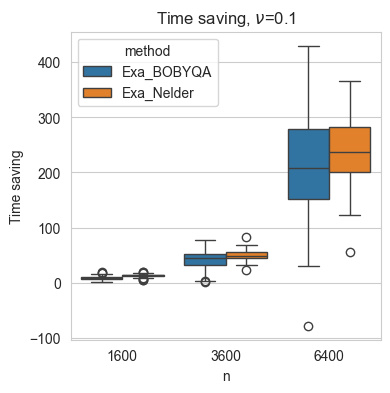}
		\caption{$\nu=0.1$}
	\end{subfigure}
	\hfill
	\begin{subfigure}[b]{0.3\textwidth}
		\centering
		\includegraphics[width=\textwidth]{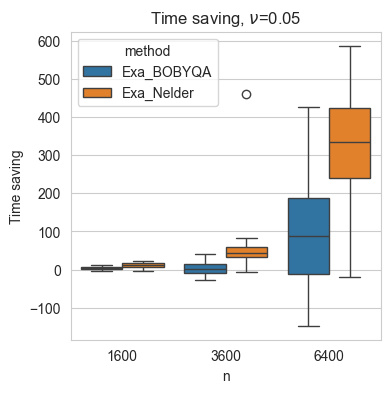}
		\caption{$\nu=0.05$}
	\end{subfigure}
	\begin{subfigure}[b]{0.3\textwidth}
		\centering
		\includegraphics[width=\textwidth]{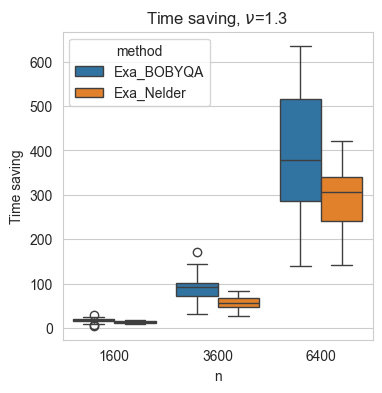}
		\caption{$\nu=1.3$}
	\end{subfigure}
	\hfill
	\caption{Boxplots of the computational time saving between the proposed method and two ExaGeoStat methods. }
	\label{fig:timediff}
\end{figure}

\begin{figure}
	\centering
	\includegraphics[width=\linewidth]{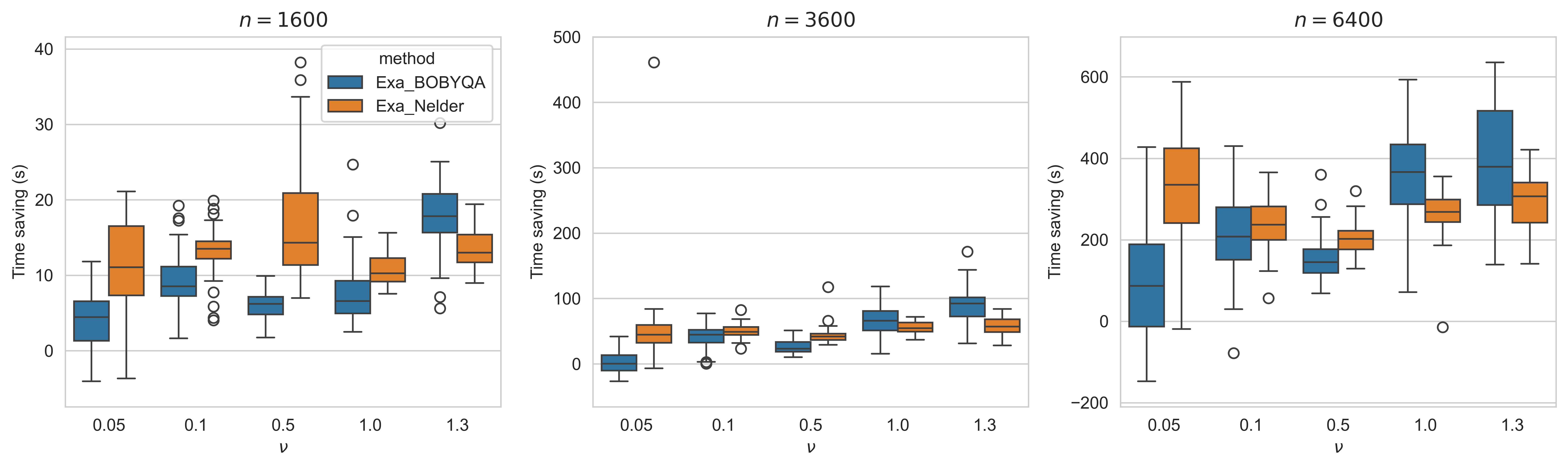}
	\caption{Boxplots of the computational time saving between the proposed method and two ExaGeoStat methods with respect to $n$. }
	\label{fig:timediff_by_n}
\end{figure}

The computational time improvement of our proposed method comes mainly from the reduced number of likelihood evaluations. 
The \texttt{ExaGeoStat}/Nelder–Mead method requires fewer likelihood function calls in most cases with larger $\nu$ values (e.g., $\nu = 1.0$ and $1.3$), whereas the  \texttt{ExaGeoStat}/BOBYQA method performs fewer calls for smaller and moderate $\nu$ values (e.g., $\nu = 0.05$, $0.1$, and $0.5$). Our proposed algorithm significantly reduces the number of likelihood function calls at the cost of increased derivative computation. For moderate and large values of $\nu$, the number of calls for derivative functions is smaller in the cases with a smaller $\nu$. When $\nu=0.05$ or $0.1$, the number of calls is close to the maximum counts of derivative calls, so the \texttt{ExaGeoStat}/Nelder-Mead algorithm is involved in this case. Even in the worst case, our proposed algorithm still has computational time comparable to that of derivative-free methods used in \texttt{ExaGeoStat}. We also show the difference in computational time between our proposed method and two \texttt{ExaGeoStat}-based methods in Figures \ref{fig:timediff}-\ref{fig:timediff_by_n}, which indicates that the proposed method has better overall computational efficiency than these two methods. Although there exist individual realizations (especially at $\nu=0.05$) in which the computational times are not optimal, our proposed approach exhibits a clear trend toward computational time savings in most replications.

\begin{figure}[htbp]
	\centering
	\begin{subfigure}[b]{0.9\textwidth}
		\centering
		\includegraphics[width=\textwidth]{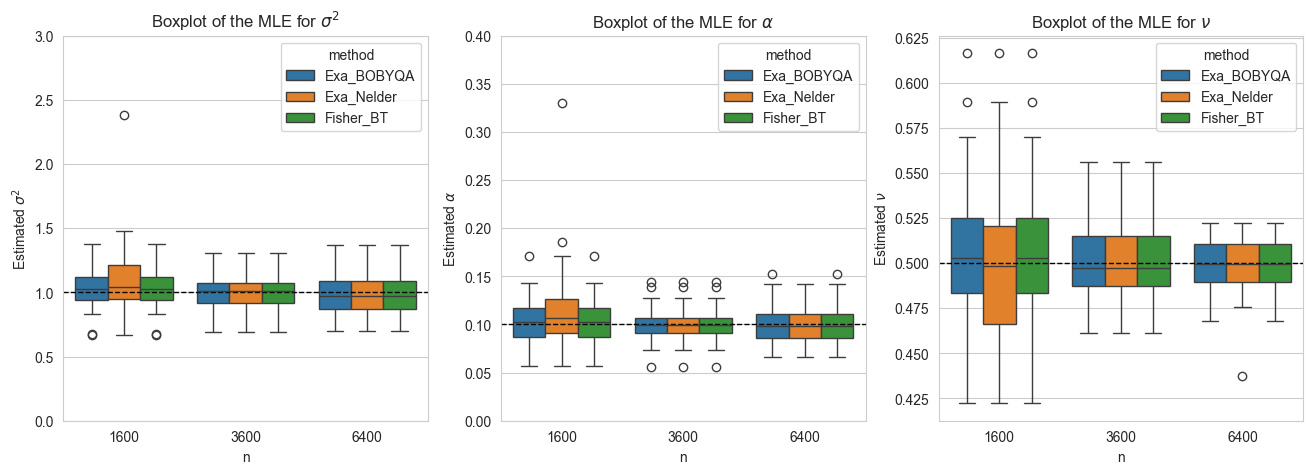}
		\caption{$(\sigma^2, \alpha, \nu) = (1.0, 0.1, 0.5)$}
		\label{subfig:boxplot_10105}
	\end{subfigure}
	\hfill
	\begin{subfigure}[b]{0.9\textwidth}
		\centering
		\includegraphics[width=\textwidth]{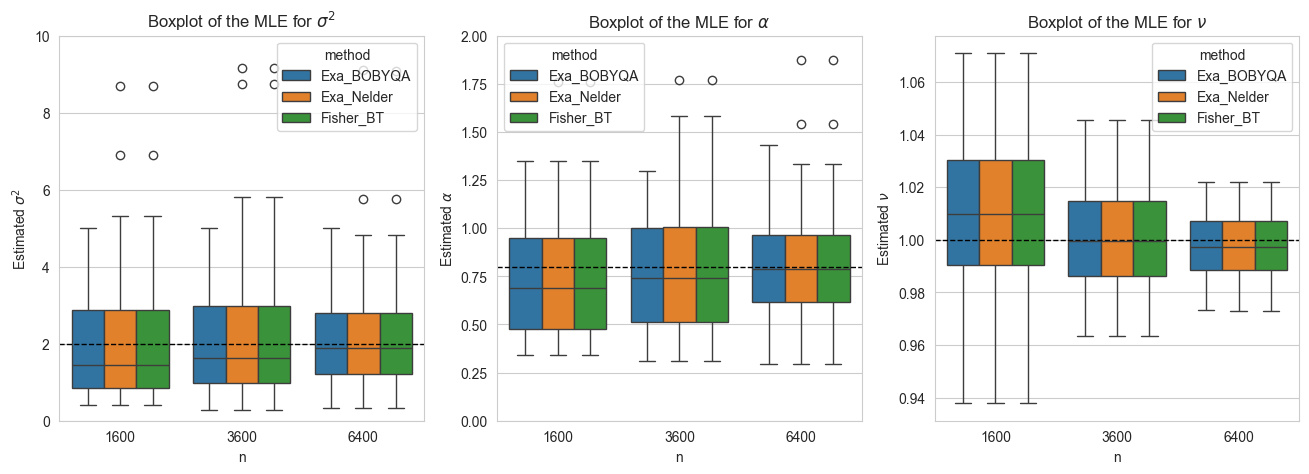}
		\caption{$(\sigma^2, \alpha, \nu) = (2.0, 0.8, 1.0)$}
		\label{subfig:boxplot_20810}
	\end{subfigure}
	\hfill
	\begin{subfigure}[b]{0.9\textwidth}
		\centering
		\includegraphics[width=\textwidth]{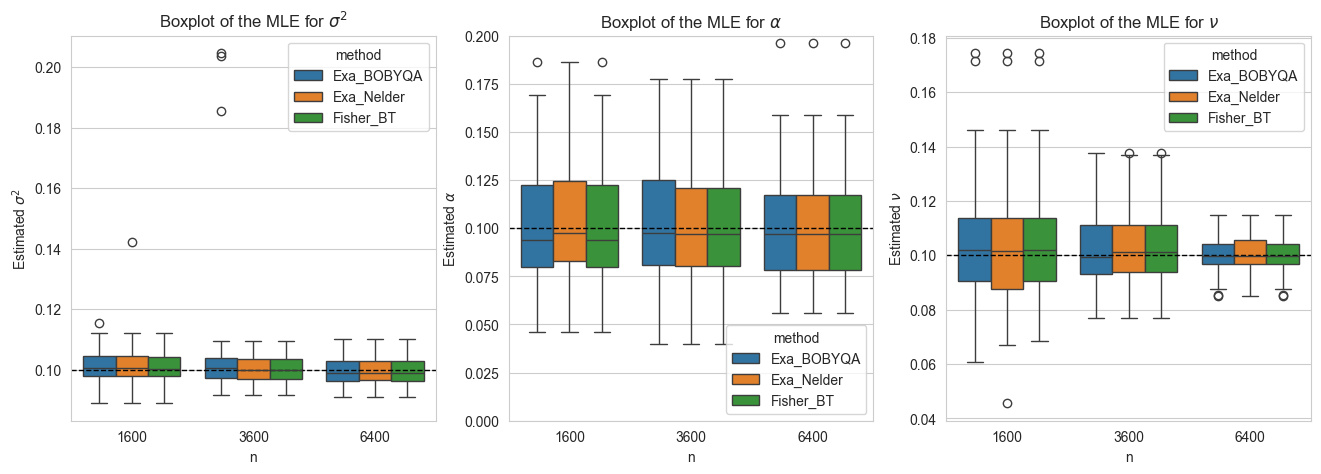}
		\caption{$(\sigma^2, \alpha, \nu) = (0.1, 0.1, 0.1)$}
		\label{subfig:boxplot_010101}
	\end{subfigure}
	\hfill
	\caption{Boxplots of estimates from different MLE algorithms for moderate true values of $\nu$. }
	\label{fig:boxplot1}
\end{figure}

\begin{figure}[htb]
	\centering
	\begin{subfigure}[b]{0.9\textwidth}
		\centering
		\includegraphics[width=\textwidth]{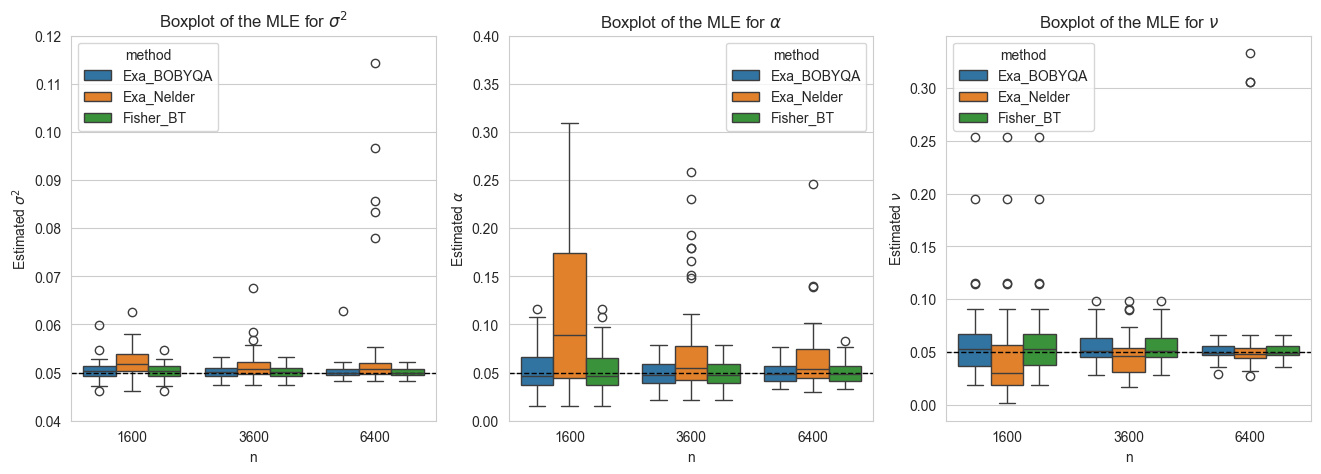}
		\caption{$(\sigma^2, \alpha, \nu) = (0.05, 0.05, 0.05)$}
		\label{subfig:boxplot_005005005}
	\end{subfigure}
	\hfill
	\begin{subfigure}[b]{0.9\textwidth}
		\centering
		\includegraphics[width=\textwidth]{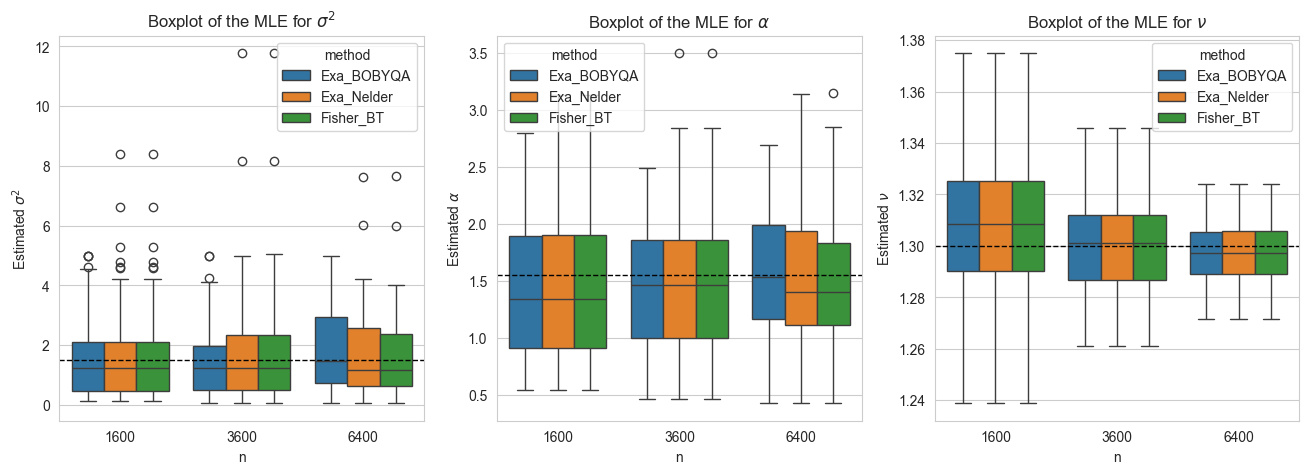}
		\caption{$(\sigma^2, \alpha, \nu) = (1.5, 1.55, 1.3)$}
		\label{subfig:boxplot_1515513}
	\end{subfigure}
	\caption{Boxplots of estimates from different MLE algorithms for small and large true values of $\nu$. }
	\label{fig:boxplot2}
\end{figure}

Next, we inspect the estimation performance of the MLE computed by three methods, as shown in Figures \ref{fig:boxplot1} and \ref{fig:boxplot2}. The \texttt{ExaGeoStat}/Nelder-Mead method may produce severe outlier estimates, especially for smaller values of $\nu$. For example, when $\sigma^2=0.5$, the largest estimate of $\sigma^2$ reaches $2.0\times 10^4$ when $n=1,600$, and $3.0\times 10^5$ when $n=6,400$. Thus, we ignore these severe outliers to make the boxplot readable. In all cases considered, the Fisher-BT algorithm yields the MLE with the smallest variance and good unbiasedness. The \texttt{ExaGeoStat}/Nelder-Mead algorithm produces boxplots similar to those of the Fisher-BT algorithm when $\nu=1.0$ and performs better than the \texttt{ExaGeoStat}/BOBYQA algorithm when $\nu=1.3$. On the other hand, the \texttt{ExaGeoStat}/BOBYQA algorithm is better when $\nu=0.05, 0.1, 0.5$, because the boxplots have a smaller variance and fewer outliers compared to their \texttt{ExaGeoStat}/Nelder-Mead counterparts. Thus, for the estimation performance, the \texttt{ExaGeoStat}/BOBYQA algorithm behaves better for small and moderate $\nu$, while the \texttt{ExaGeoStat}/Nelder-Mead algorithm is better for large $\nu$. However, our Fisher-BT algorithm is capable of giving estimates with good precision for all considered $\nu$, so our proposed algorithm also has the advantage of allowing for a wider range of the true value of $\nu$. 


\begin{figure}[htb]
	\centering
	\begin{subfigure}[b]{0.3\textwidth}
		\centering
		\includegraphics[width=\textwidth]{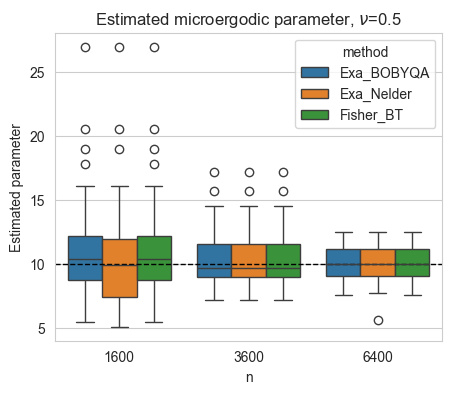}
		\caption{$\nu=0.5$}
		\label{subfig:boxmicro_10105}
	\end{subfigure}
	\hfill
	\begin{subfigure}[b]{0.3\textwidth}
		\centering
		\includegraphics[width=\textwidth]{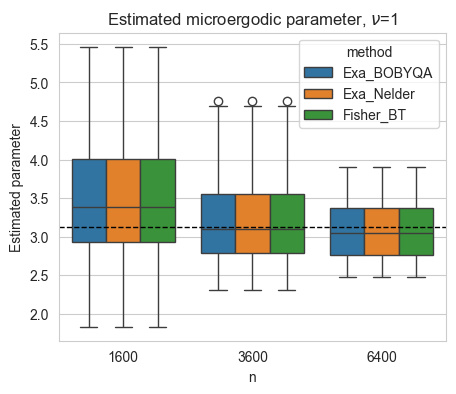}
		\caption{$\nu=1.0$}
		\label{subfig:boxmicro_20810}
	\end{subfigure}
	\hfill
	\begin{subfigure}[b]{0.3\textwidth}
		\centering
		\includegraphics[width=\textwidth]{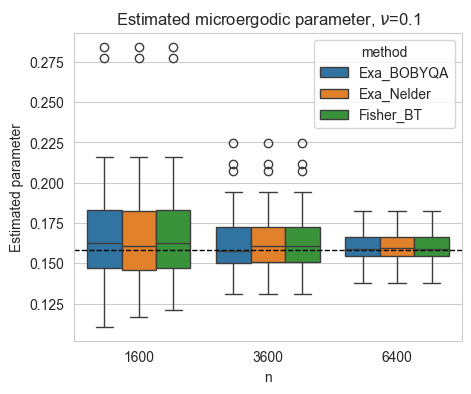}
		\caption{$\nu=0.1$}
		\label{subfig:boxmicro_010101}
	\end{subfigure}
	\hfill
	\begin{subfigure}[b]{0.3\textwidth}
		\centering
		\includegraphics[width=\textwidth]{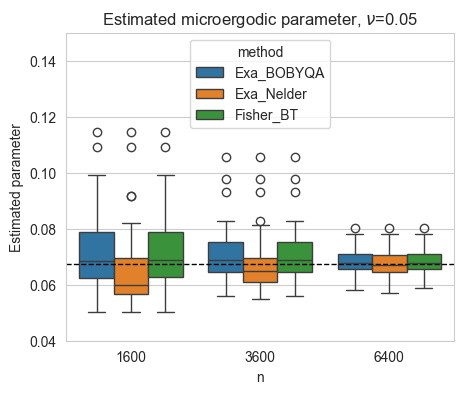}
		\caption{$\nu=0.05$}
		\label{subfig:boxmicro_005005005}
	\end{subfigure}
	\begin{subfigure}[b]{0.3\textwidth}
		\centering
		\includegraphics[width=\textwidth]{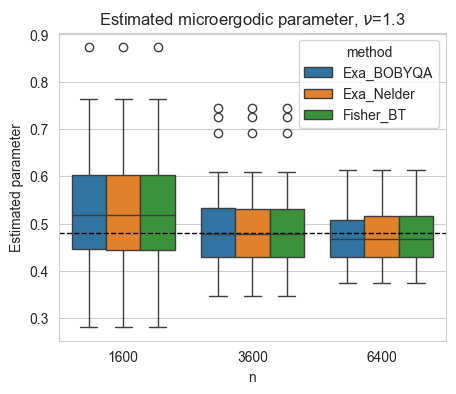}
		\caption{$\nu=1.3$}
		\label{subfig:boxmicro_1515513}
	\end{subfigure}
	\hfill
	\caption{Boxplots of the microergodic parameter estimates from different MLE algorithms. }
	\label{fig:boxmicro}
\end{figure}

In Figures \ref{fig:boxplot1} and~\ref{fig:boxplot2}, the estimation variance of $\nu$ decreases when $n$ increases, but this variance for $\sigma^2$ and $\alpha$ does not have a significant change with different $n$. Since we do not change the size of the observation area and increase the density of observation locations, which corresponds to a fixed domain asymptotic framework, the parameters $\sigma^2$ and $\alpha$ cannot be consistently estimated according to \cite{zhang2004inconsistent}. However, \cite{zhang2004inconsistent} pointed out that the microergodic parameter, $\theta_m = \sigma^2 \alpha^{-2\nu}$, can be consistently estimated under fixed domain asymptotics. Thus, we also show the boxplot of the MLE for $\theta_m$ under different algorithms, which is shown in Figure \ref{fig:boxmicro}. Note that we removed six outliers only in the case of $\nu = 0.05$ to improve the clarity of the plot. Figure \ref{fig:boxmicro} shows that the MLE for $\theta_m$ has a smaller variance with larger $n$, which is similar to the asymptotic result for fixed $\nu$ in \cite{zhang2004inconsistent}. The \texttt{ExaGeoStat}/Nelder-Mead algorithm performs worse than the other two algorithms when $\nu=0.05$ or $\nu=0.5$ and $n=1,600$, whereas our considered methods provide similar estimates of $\theta_m$ in the other cases. 

\begin{figure}[htbp]
	\centering
	\begin{subfigure}[b]{0.3\textwidth}
		\centering
		\includegraphics[width=\textwidth]{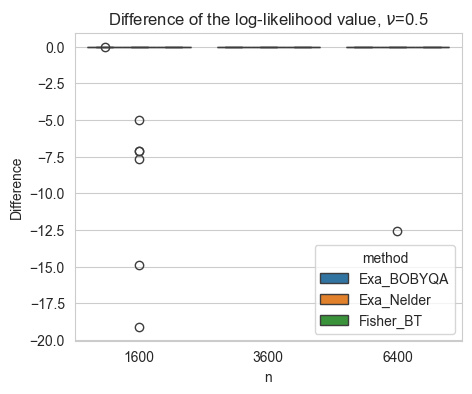}
		\caption{$\nu=0.5$}
		\label{subfig:loglik_10105}
	\end{subfigure}
	\hfill
	\begin{subfigure}[b]{0.3\textwidth}
		\centering
		\includegraphics[width=\textwidth]{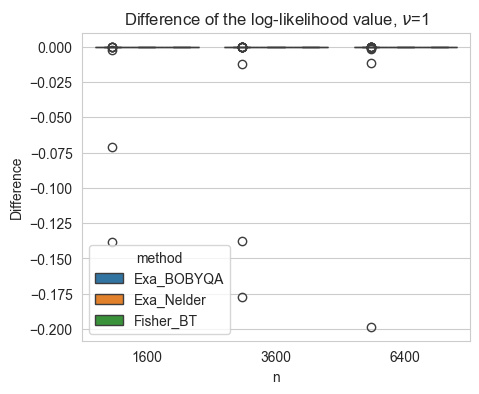}
		\caption{$\nu=1.0$}
		\label{subfig:loglik_20810}
	\end{subfigure}
	\hfill
	\begin{subfigure}[b]{0.3\textwidth}
		\centering
		\includegraphics[width=\textwidth]{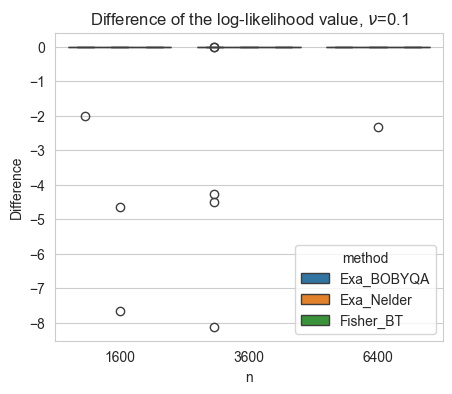}
		\caption{$\nu=0.1$}
		\label{subfig:loglik_010101}
	\end{subfigure}
	\hfill
	\begin{subfigure}[b]{0.3\textwidth}
		\centering
		\includegraphics[width=\textwidth]{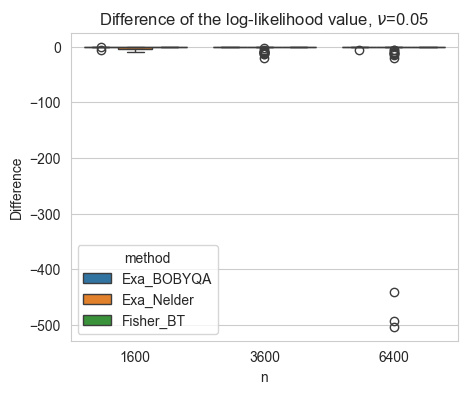}
		\caption{$\nu=0.05$}
		\label{subfig:loglik_005005005}
	\end{subfigure}
	\begin{subfigure}[b]{0.3\textwidth}
		\centering
		\includegraphics[width=\textwidth]{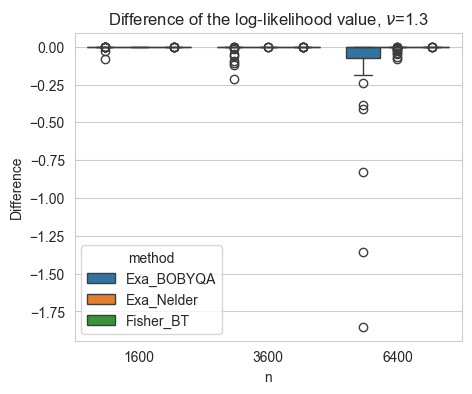}
		\caption{$\nu=1.3$}
		\label{subfig:loglik_1515513}
	\end{subfigure}
	\hfill
	\caption{Boxplots of the difference between the computed log-likelihood values and their maximum values. }
	\label{fig:loglik}
\end{figure}

Finally, to evaluate the efficiency of the optimization algorithm, we compare the difference between the computed log-likelihood value of each optimization algorithm and the maximum log-likelihood value for the same data among three MLE optimization methods, the result of which is shown in Figure \ref{fig:loglik}. We ignore the case where the absolute value of this difference is not greater than $10^{-6}$. Figure \ref{fig:loglik} shows that, when $\nu$ is small or moderate, such as $\nu=0.05, 0.1, 0.5$, the \texttt{ExaGeoStat}/Nelder-Mead method is more likely to produce a suboptimal log-likelihood value. However, when $\nu$ is larger, such as $\nu=1.0, 1.3$, the \texttt{ExaGeoStat}/BOBYQA method can be suboptimal for more cases. In most cases, the proposed Fisher-BT method yields optimal results among the three methods. Even in some cases where the Fisher-BT method is suboptimal, which appears when $\nu=1.3$, the loss of the log-likelihood function is smaller compared to the other two methods since the largest log-likelihood value loss is $5.4\times 10^{-5}$. Thus, our proposed method also outperforms the \texttt{ExaGeoStat}/BOBYQA and \texttt{ExaGeoStat}/Nelder-Mead methods in this simulation.

\section{Application to Soil Moisture Data} \label{Sec:applications}

\begin{figure}[htbp]
	\centering
	\includegraphics[width=0.5\textwidth]{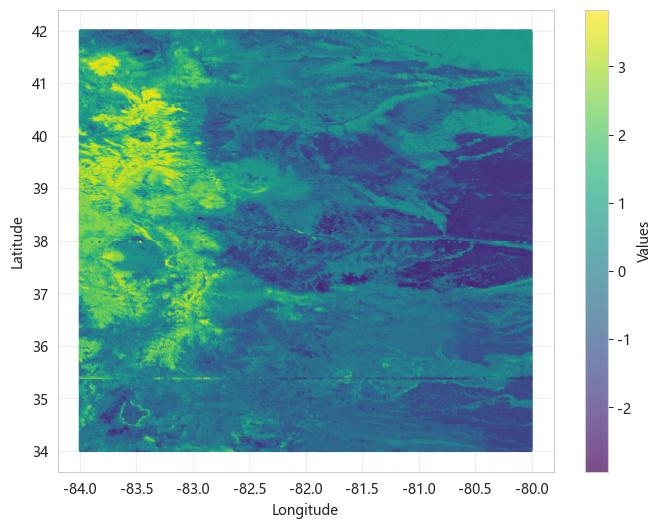}
	\caption{Illustration of the soil moisture dataset in our real-case study. }
	\label{fig:soil}
\end{figure}

We consider the soil moisture dataset introduced by~\cite{huang2018hierarchical}. The dataset consists of high-resolution daily soil moisture data for the top layer of the Mississippi Basin in the US, as of January 1, 2004. \cite{huang2018hierarchical} fitted this data by a linear model with the longitude and latitude as covariates, and then used a logarithmic transformation with some shift to the residuals, obtaining a Gaussian-distributed soil moisture residual dataset. We truncate the data within a range of $[-84^\circ E, -80^\circ E]\times [34^\circ N, 42^\circ N]$, resulting in a dataset with $n_{\text{all}}=542,629$ observations. An illustration of the soil moisture dataset considered in this work is shown in Figure \ref{fig:soil}. 

\begin{figure}[htbp]
	\centering
	\begin{subfigure}[b]{0.7\textwidth}
		\centering
		\includegraphics[width=\textwidth]{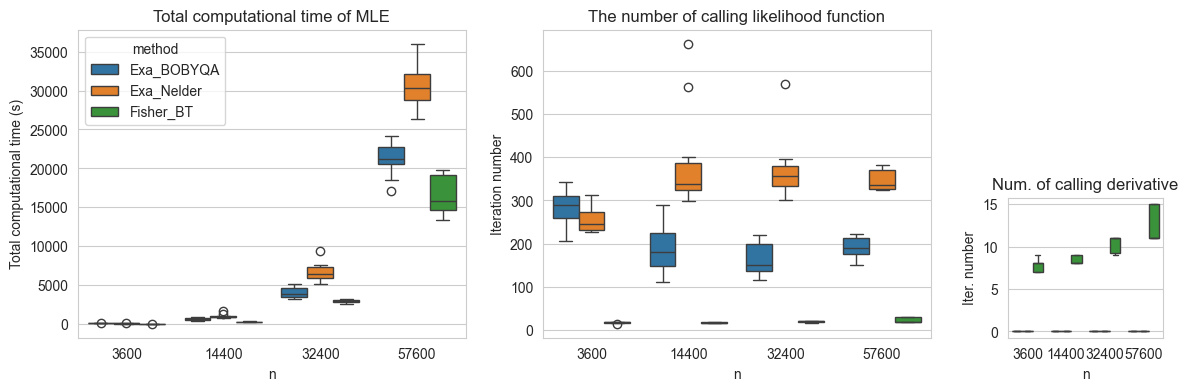}
	\end{subfigure}
	\begin{subfigure}[b]{0.25\textwidth}
		\centering
		\includegraphics[width=\textwidth]{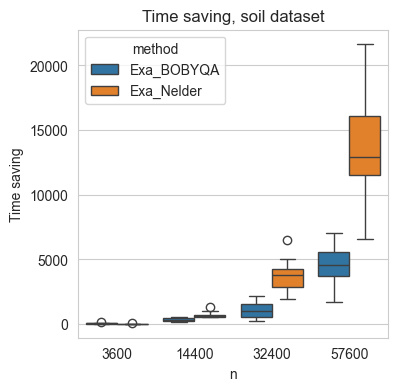}
	\end{subfigure}
	\caption{Boxplots of the computational time, the number of iterations, and the computational time difference for soil moisture dataset estimation. }
	\label{fig:real_res}
\end{figure}

To compare computational accuracy and estimation performance across different sample sizes, we consider subsets of sizes $n=3,600$, $14,400$, $32,400$, and $57,600$. For each $n$, we uniformly draw 10 random subsamples of size $n$ and then evaluate the computational time, the number of iterations, and the estimation results. We choose $(\sigma^2, \alpha, \nu) \in [0.01, 5]\times [0.01, 5] \times [0.01, 2]$ as the searching range of the \texttt{ExaGeoStat}/BOBYQA method, setting the lower extreme $(0.01, 0.01, 0.01)$ and upper extreme $(5, 5, 2)$ as the initial guess vectors in our proposed method and the middle point $(2.505, 2.505, 1.005)$ as the initial value of the \texttt{ExaGeoStat}/Nelder-Mead method. Figure \ref{fig:real_res} shows the computational time, the number of iterations, and the computational time difference between the proposed method and other methods. Table \ref{tab:real_est} shows the mean and standard deviation of the estimated parameters for different $n$. 

\begin{table}[htb]
	\centering
	\caption{Mean (Standard Deviation) of the estimates for different sample sizes and methods in the soil moisture dataset estimation}
	\label{tab:real_est}
	\resizebox{\textwidth}{!}{
		\begin{tabular}{lccccccccc}
			\toprule
			\multirow{2}{*}{$n$} & \multicolumn{3}{c}{$\sigma^2$} & \multicolumn{3}{c}{$\alpha$} &  \multicolumn{3}{c}{$\nu$} \\
			& Exa\_BOBYQA & Exa\_Nelder & Fisher\_BT & Exa\_BOBYQA & Exa\_Nelder & Fisher\_BT & Exa\_BOBYQA & Exa\_Nelder & Fisher\_BT \\
			\midrule
			\multirow{2}{*}{3,600} & 1.5213 & 1.5214 & 1.5214 & 2.8993 & 2.8997 & 2.8997 & 0.2420 & 0.2420 & 0.2420 \\ 
			& (0.0488) & (0.0488) & (0.0488) & (0.3610) & (0.3612) & (0.3612) & (0.0108) & (0.0108) & (0.0108) \\ 
			\multirow{2}{*}{14,400} & 1.3366 & 1.3366 & 1.3366 & 1.3406 & 1.3406 & 1.3406 & 0.2773 & 0.2773 & 0.2773 \\ 
			& (0.0394) & (0.0394) & (0.0394) & (0.1307) & (0.1307) & (0.1307) & (0.0050) & (0.0050) & (0.0050) \\ 
			\multirow{2}{*}{32,400} & 1.2122 & 1.2122 & 1.2122 & 0.7553 & 0.7553 & 0.7553 & 0.3114 & 0.3114 & 0.3114 \\ 
			& (0.0128) & (0.0127) & (0.0127) & (0.0413) & (0.0413) & (0.0413) & (0.0053) & (0.0053) & (0.0053) \\ 
			\multirow{2}{*}{57,600} & 1.1307 & 1.1307 & 1.1307 & 0.5045 & 0.5045 & 0.5045 & 0.3397 & 0.3397 & 0.3397 \\ 
			& (0.0130) & (0.0130) & (0.0130) & (0.0321) & (0.0322) & (0.0322) & (0.0061) & (0.0061) & (0.0061) \\ 
			\bottomrule
		\end{tabular}
	}
\end{table}


As shown in Figure \ref{fig:soil}, the proposed Fisher-BT method reduces the computational time in all considered cases. Although the proposed method requires computing the derivative of the log-likelihood function, it still benefits from a significantly smaller number of log-likelihood function evaluations. Table \ref{tab:real_est} shows that the methods we considered yield similar estimates across all cases. In fact, the estimates for parameters  $\sigma^2, \alpha, \nu$ across different methods are nearly identical, with maximum differences of only $3.34\times 10^{-4}$, $1.70\times 10^{-3}$, and $2.62\times 10^{-6}$, respectively. 
The proposed method achieves the largest log-likelihood value across all datasets, except for two samples with $n=57,600$, where its value is merely $1.35\times 10^{-10}$ and $3.64\times 10^{-12}$ smaller than the \texttt{ExaGeoStat}/BOBYQA method, respectively. 
In conclusion, for the soil moisture subsample dataset considered in this study, our proposed method can significantly reduce computational time while achieving competitive estimation accuracy.

\section{Conclusions and Discussions} \label{Sec:conclusions}

This work proposes the Fisher-BT optimization algorithm for evaluating the exact MLE of the stationary Mat\'{e}rn covariance model for large spatial datasets. We introduce a two-stage structure into this algorithm by combining Fisher scoring with a backtracking line search and the Nelder-Mead method. 
The Fisher scoring step accelerates convergence by reducing the number of calls of the log-likelihood function, for which the convergence is stabilized by a simplification of \cite{geoga2023fitting}'s series approximation method. 
The Nelder-Mead step serves as a safeguard when Fisher scoring fails to converge, yielding more robust estimates over a broader range of values of the smoothness parameter $\nu$.
With the \texttt{ExaGeoStat} framework, our method can be implemented on multi- and many-core systems.  
Simulations and real-data applications show that our proposed method reduces the computational time of the exact MLE by minimizing the number of calls to the log-likelihood functions while maintaining estimation accuracy. Simulations also show that our proposed method is more robust to smaller or larger values of $\nu$ than the \texttt{ExaGeoStat}/BOBYQA and \texttt{ExaGeoStat}/Nelder-Mead methods. 

Our proposed algorithm can be easily generalized to other stationary covariance models, such as those incorporating a nugget term or geometric anisotropy. In future work, we plan to combine our algorithm with matrix approximation techniques, such as the tile low-rank approximation \citep{abdulah2018exageostat}, thereby developing a more computationally efficient method suitable for very large spatial datasets with $n>10^5$. 

The primary obstacle to the robustness of our method lies in the computational accuracy of the modified Bessel function of the second kind. Computing the second-order derivative is less robust and consumes more computational time, so we did not adopt the Newton method for optimization. Thus, our proposed method can benefit significantly, or even eliminate the Nelder-Mead step, if a more numerically stable algorithm can be found to compute the ${\cal K}_\nu$ function and its derivatives. Our proposed method can also be used with the unbiased estimation equation method proposed by \cite{sun2016statistically}, thereby relaxing the requirement of a fixed smoothness parameter due to computational instability. In summary, our proposed Fisher-BT algorithm provides a fast, accurate, and reliable tool for exact MLE computation and establishes a benchmark for assessing the computational accuracy of MLE approximation methods.

\section{Disclosure statement}\label{disclosure-statement}

The authors has no conflicts of interest exist. 

\section{Data Availability Statement}\label{data-availability-statement}

Deidentified data have been made available in the Supplementary Material C. 

\phantomsection\label{supplementary-material}
\bigskip

\begin{center}

{\large\bf SUPPLEMENTARY MATERIAL}

\end{center}

\begin{description}
	
\item[A. Algorithm for computing partial derivatives:]
A document presents the detailed algorithm for computing $\symbf{\Sigma}_\nu$, the partial derivative of the covariance matrix with respect to the smoothness parameter.

In this Supplementary Material A, we introduce the algorithm for computing $\pmb{\Sigma}_\nu$ based on the series approximation algorithm proposed by \cite{geoga2023fitting}. 
In the program, the conditions for computing the derivatives are: 
\begin{enumerate}
	\item Check if $x=0$. The derivative is zero when $x=0$; 
	\item If not, check $x < 8.5$ and $|\nu - \left[ \nu \right]| > 10^{-6}$, where $\left[ \nu \right]$ is the nearest integer of $\nu$; 
	\item If not, check if $x \ge 8.5$ and $x < 30$; 
	\item If not, check if $x \ge 30$ and $|(\nu + 0.5) - \left[ \nu + 0.5 \right]| > 10^{-6}$; 
	\item If not, using difference approximation with lag $10^{-9}$ to compute the derivative. 
\end{enumerate}

Here are the algorithms for cases 2-4: 

\textbf{Case 2: $0 < x < 8.5$, $\nu$ is not close to an integer. }

In this case, the derivative of $x^\nu {\cal K}_\nu(x)$, where ${\cal K}_\nu$ is the modified Bessel function of the second kind, satisfies 
\begin{equation*}
	{\cal K}_\nu(x) = \sum_{k=0}^\infty \left( \frac{x}{2} \right)^{2k} \frac{1}{2 \cdot k!} \left\{ \Gamma(\nu) \left( \frac{x}{2} \right)^{-\nu} \frac{\Gamma(1 - \nu)}{\Gamma(1 + k - \nu)} + \Gamma(-\nu) \left( \frac{x}{2} \right)^{\nu} \frac{\Gamma(1 + \nu)}{\Gamma(1 + k + \nu)} \right\}. 
\end{equation*}
Let $g_k(\nu) = \Gamma(1 + \nu) / \Gamma(1 + k + \nu)$, then 
\begin{equation*}
	g_k(\nu) = \begin{cases}
		\prod_{j=1}^k (\nu + j)^{-1}, & k \ge 1, \\
		1, & k = 0. \\
	\end{cases}
	\quad 
	d_k(\nu) := \frac{\partial}{\partial \nu} \log(g_k(\nu)) = \begin{cases}
		-\sum_{j=1}^k (\nu + j)^{-1}, & k \ge 1, \\
		0, & k = 0. \\
	\end{cases}
\end{equation*}

By direct computation, 
\begin{equation*}
	\begin{aligned}
		\frac{\partial}{\partial \nu} [x^\nu {\cal K}_\nu(x)] & = \sum_{k=0}^\infty \left( \frac{x}{2} \right)^{2k} \frac{1}{2 \cdot k!} \Big[ 2^{\nu} \left\{ \log 2 + \psi(\nu) - d_k(-\nu) \right\} \Gamma(\nu) g_k(-\nu) \\ 
		& + 2^{-\nu} x^{2\nu} \left\{ -\log 2 + 2\log (x) - \psi(-\nu) + d_k(\nu) \right\} \Gamma(-\nu) g_k(\nu) \Big], 
	\end{aligned}
\end{equation*}
where $\psi(\nu) = \Gamma^\prime (\nu) / \Gamma(\nu)$ is the digamma function. 
The series is approximated by the partial sum up to the 20th term. 

\textbf{Case 3: $8.5 \le x < 30$. }

When $\nu \to \infty$, ${\cal K}_\nu (\nu x)$ is approximated by 
\begin{equation*}
	{\cal K}_\nu (\nu x) \approx \sqrt{\frac{\pi}{2 \nu}} \frac{\exp(-\nu \eta(x))}{(1 + x^2)^{1/4}} \sum_{k=0}^\infty (-1)^k \nu^{-k} U_k(p(x)), 
\end{equation*}
where 
\begin{equation*}
	\eta(x) = \sqrt{1 + x^2} + \log \left( \frac{x}{1 + \sqrt{1 + x^2}} \right), \quad p(x) = (1+x^2)^{-1/2}, 
\end{equation*}
and $U_k(p)$ is the polynomial of $3k$ order, defined recursively by $U_0(p) \equiv 1$, 
\begin{equation*}
	U_{k+1}(p) = \frac{1}{2}p^2(1-p^2) U_k^\prime (p) + \frac{1}{8} \int_0^p (1 - 5t^2) U_k(t) dt. 
\end{equation*}
Let $U_k(p) = \sum_{j=0}^{3k} c_j^{(k)} p^j$, then by direct computation, 
\begin{equation*}
	\begin{aligned}
		& c_0^{(k+1)} = 0; \quad c_1^{(k+1)} = c_0^{(k)} / 8; \quad c_2^{(k+1)} = 9c_1^{(k)} / 16; \quad c_3^{(k+1)} = 25c_2^{(k)} / 24 - 5c_0^{(k)} / 24; \\
		& c_j^{(k+1)} = \frac{1}{2} \left( j - 1 + \frac{1}{4j} \right) c_{j-1}^{(k)} - \frac{1}{2} \left( j - 3 + \frac{5}{4j} \right) c_{j-3}^{(k)}, \quad j \in \{4, \ldots, 3k+1\}; \\
		& c_j^{(k+1)} = -\frac{1}{2} \left( j - 3 + \frac{5}{4j} \right) c_{j-3}^{(k)}, \quad j \in \{3k+2, 3k+3\}. \\
	\end{aligned}
\end{equation*}
By direct computation, 
\begin{equation*}
	\begin{aligned}
		& \frac{\partial}{\partial \nu} [x^\nu {\cal K}_\nu(x)] \\ 
		&\approx \left[ \log \nu + \log \left\{ 1 + \sqrt{1 + \left(\frac{x}{\nu}\right)^2} \right\} - \frac{1}{2\nu \left\{ 1 + (x/\nu)^2 \right\}} \right] g_\nu(x) \sum_{k=0}^\infty (-1)^k \nu^{-k} U_k(p(x/\nu)) \\ 
		&- g_\nu(x) \sum_{k=0}^\infty (-1)^{k} k \nu^{-k-1} U_k(p(x/\nu)) \\
		&+ \frac{x^2}{\nu^3} \left\{ 1 + (x/\nu)^2 \right\} ^{-3/2} g_\nu(x) \sum_{k=0}^\infty (-1)^k \nu^{-k} U_k^\prime (p(x/\nu)), 
	\end{aligned}
\end{equation*}
where 
\begin{equation*}
	g_\nu(x) = x^\nu \sqrt{\frac{\pi}{2\nu}} \frac{\exp(-\nu \eta(x/\nu))}{\{1 + (x/\nu)^2\}^{1/4}}. 
\end{equation*}
The series is approximated by the partial sum up to the 12-th term when $8.5 \le x < 15$ and up to the 8-th term when $15 \le x < 30$. 

\textbf{Case 4: $x \ge 30$ and $\nu+0.5$ is not close to integer. }  \label{Sec:x_ge_30}

When $x \to \infty$, ${\cal K}_\nu(x)$ is approximated by 
\begin{equation*} 
	{\cal K}_\nu(x) = \sqrt{\frac{\pi}{2x}} e^{-x} \sum_{k=0}^\infty x^{-k} a_k(\nu), 
\end{equation*}
where 
\begin{equation*}
	a_k(\nu) = \begin{cases}
		1, & k = 0; \\
		\frac{1}{8^k\Gamma(k+1)} \prod_{j=1}^k \{ 4\nu^2 - (2j-1)^2 \}, & k \ge 1. 
	\end{cases}
\end{equation*}
By direct computation, 
\begin{equation*}
	\frac{\partial}{\partial \nu} [x^\nu {\cal K}_\nu(x)] \approx \sqrt{\frac{\pi}{2}} x^{\nu - 1/2} e^{-x} \sum_{k=0}^\infty x^{-k} \{ \log x + b_k(\nu) \} a_k(\nu), 
\end{equation*}
where 
\begin{equation*}
	b_k(\nu) = \begin{cases}
		0, & k=0; \\
		\sum_{j=1}^k \frac{8\nu}{4\nu^2 - (2j-1)^2}, & k \ge 1. 
	\end{cases}
\end{equation*}
The series is approximated by the partial sum up to the 5th term when $x \ge 30$. 

\item[B. Code:]
A ZIP file containing C and Python codes for reproducing the simulation and soil moisture application results. 
\item[C. Soil moisture dataset:]
A ZIP file containing the soil moisture dataset used in the article, along with the code for subsampling. 
\end{description}

  \bibliography{bibliography.bib}

\end{document}